\documentclass[lettersize,journal]{IEEEtran}
\usepackage{amsmath,amsfonts}
\usepackage{algorithmic}
\usepackage{algorithm}
\usepackage{array}

\usepackage{xcolor}
\usepackage{subcaption}

\definecolor{lightblue}{RGB}{0, 122, 204} 

\usepackage[
    colorlinks=true,
    linkcolor=lightblue,
    citecolor=lightblue,
    urlcolor=lightblue
]{hyperref}
\usepackage{textcomp}
\usepackage{stfloats}
\usepackage{url}
\usepackage{verbatim}
\usepackage{graphicx}
\begin{document}

\title{Teaching Cars to Drive: Spotlight on Connected and Automated Vehicles}

\author{Filippos N. Tzortzoglou, \IEEEmembership{Student Member, IEEE,} and Andreas A. Malikopoulos,
\IEEEmembership{Senior Member, IEEE}
\thanks{This research was supported in part by NSF under Grants CNS-2401007, CMMI-2348381, IIS-2415478, and in part by MathWorks.} \thanks{Filippos N. Tzortzoglou, and Andreas A. Malikopoulos are with the Department of Civil and Environmental Engineering, Cornell University, Ithaca, NY 14853 USA.   (emails: \tt\small{ft253@cornell.edu; amaliko@cornell.edu)}}}



\maketitle

In recent decades, society has witnessed significant advancements in emerging mobility systems. These systems refer to transportation solutions that incorporate digital technologies, automation, connectivity, and sustainability to create safer, more efficient, and user-centered mobility. Examples include connected and automated vehicles (CAVs), shared mobility services (car-pooling), electric vehicles, and mobility-as-a-service platforms. These innovations have the potential to greatly impact areas such as safety, pollution, comfort, travel time, and fairness. In this article, we explore the current landscape of CAVs. We discuss their role in daily life and their future potential, while also addressing the challenges they may introduce. Following, we also examine the practical difficulties in research associated with CAVs especially simulating and testing CAV-related algorithms in real-world settings. We present existing solutions that aim to overcome these limitations. Finally, we provide an accessible introduction to modeling CAVs using basic kinematic principles and offer an open-source tutorial to help interested students begin exploring the field.

\vspace{-4pt}
\section{Introduction}
\begin{figure*}
    \centering
    \includegraphics[width=0.99\linewidth]{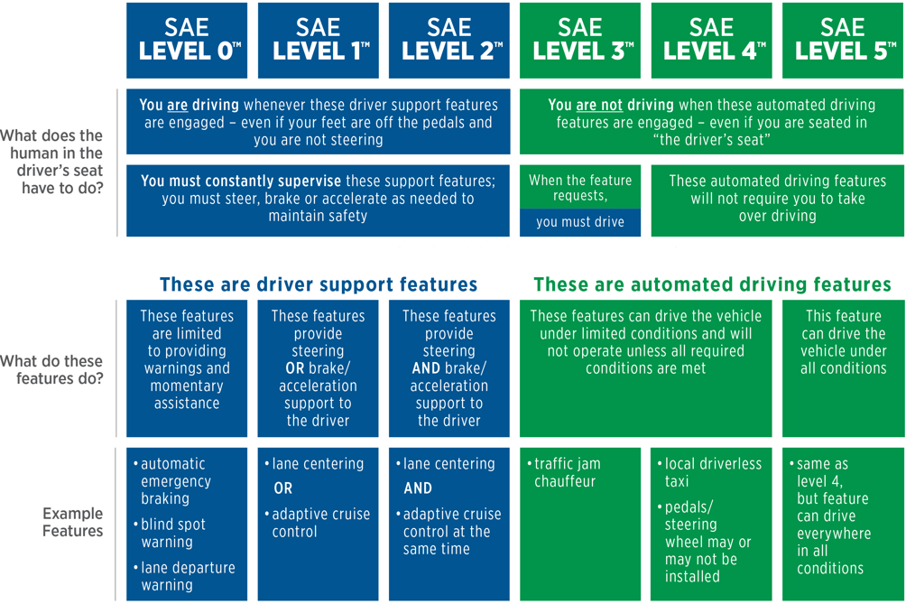}
    \caption{Levels of automation based on Society of Automotive Engineers.}
    \label{fig:SAe LEVELS}
    \vspace{-12pt}
\end{figure*}
\IEEEPARstart{I}{magine} a commuter setting out on a typical morning drive in a commercial human-driven vehicle (HDV). What seems like a routine trip often involves hidden risks such as safety concerns, high levels of emissions, discomfort, unexpected delays, or unfairness in the road network. We begin our exposition by exploring the significance of each of these issues and how they can negatively impact the performance and fairness of modern transportation.

Regarding safety, human error remains a leading cause of road incidents, including delayed reaction times, driver distraction, and misinterpretation of pedestrian behavior. In the United States alone, human error is responsible for approximately 94\% of all car crashes according to the National Highway Traffic Safety Administration (NHTSA). In 2023, there were around 40.990 traffic-related fatalities, only slightly down from the previous year and still higher than any pre-pandemic year since 2008. On a global scale, road crashes cause an estimated 1.19 million deaths and 20–50 million injuries each year.

Additionally, energy consumption and environmental pollution are also noticeable concerns of modern transportation. Unlike automated systems, human drivers typically lack real-time awareness of how specific actions, such as unnecessary acceleration or inefficient routing, affect fuel/battery efficiency. Similarly, travel time is often suboptimal due to drivers’ limited knowledge of traffic conditions leading to increased energy consumption. According to the Texas A\&M Transportation Institute, in 2022, traffic congestion in the United States caused commuters in large urban areas to waste approximately 295 million hours, burn billions of gallons of fuel, and cost the trucking industry alone over \$108 billion in delays and inefficiencies.

In terms of comfort, prolonged driving can result in fatigue, reduced alertness, and physical discomfort. Only fatigue is involved in over 91,000 crashes annually according to NHTSA. For this reason, many transportation authorities recommend taking breaks every two hours of continuous driving. Beyond fatigue, the manual demands of operating an HDV, such as gear shifting, monitoring traffic signals, interpreting GPS directions, and reacting to real-time conditions, contribute to a less seamless and more stressful experience.

Finally, fairness and accessibility remain overlooked challenges. In a human-driven transportation network, coordination between vehicles is limited or absent, making it difficult to ensure equitable access to essential services such as hospitals. Several factors, including the urgency of the trip, or the distance between essential services and users are rarely considered in current traffic systems.

As automated vehicles (AVs) continue penetrating the market, many of the critical issues discussed earlier will gradually fade from our daily transportation experience. AVs refer to vehicles equipped with varying degrees of automation, as categorized by the Society of Automotive Engineers (SAE) levels ranging from level 0 (warnings or momentary assistance) to level 5 (full automation without human input); see Fig. \ref{fig:SAe LEVELS}.

One of the main features that makes AVs, with higher SAE level of autonomy, appealing is their ability to take over actions from the driver and perform them autonomously using sophisticated decision-making algorithms. According to the level of autonomy, these algorithms are designed to guarantee safety and minimize objectives such as energy consumption or/and travel time, while simultaneously maximizing overall traffic flow. For example, if an AV is classified as SAE level 5, a human is no longer required to manage the gas pedal, brake, or steering wheel. Instead, intelligent control systems take over these functions, making decisions from both a high-level perspective, such as selecting routes based on traffic conditions, and a low-level perspective, such as computing an optimal acceleration profile to pass through a traffic light.

When AVs are equipped with communication capabilities--for example, to exchange information with smart traffic lights for more efficient crossings--they are referred to as connected and automated vehicles (CAVs). Note, that CAVs do not require full automation (SAE level 5); rather, a CAV may operate at any SAE level (from 0 to 5) as long as it integrates communication technologies. The connectivity of CAVs can take several forms:
\begin{itemize}
\item \textbf{Vehicle-to-Infrastructure (V2I)} communication, where vehicles exchange information with traffic lights or road sensors.
\item \textbf{Vehicle-to-Vehicle (V2V)} communication, enabling CAVs to coordinate with other vehicles.
\item \textbf{Vehicle-to-Everything (V2X)} communication, a broader framework that includes both V2I and V2V, enabling comprehensive situational awareness.
\end{itemize}

Together, autonomy and connectivity form the foundation of the CAV ecosystem, enabling smarter, safer, and more sustainable transportation systems.
The rest of this article explores current and emerging technologies of the domain while discussing challenges associated with research of CAVs, and practical methods to address them. We also provide a simple modeling example and an open-source tutorial to help readers start simulating CAVs.

\begin{figure*}
    \centering
    \begin{subfigure}[b]{0.48\textwidth}
        \includegraphics[width=\textwidth]{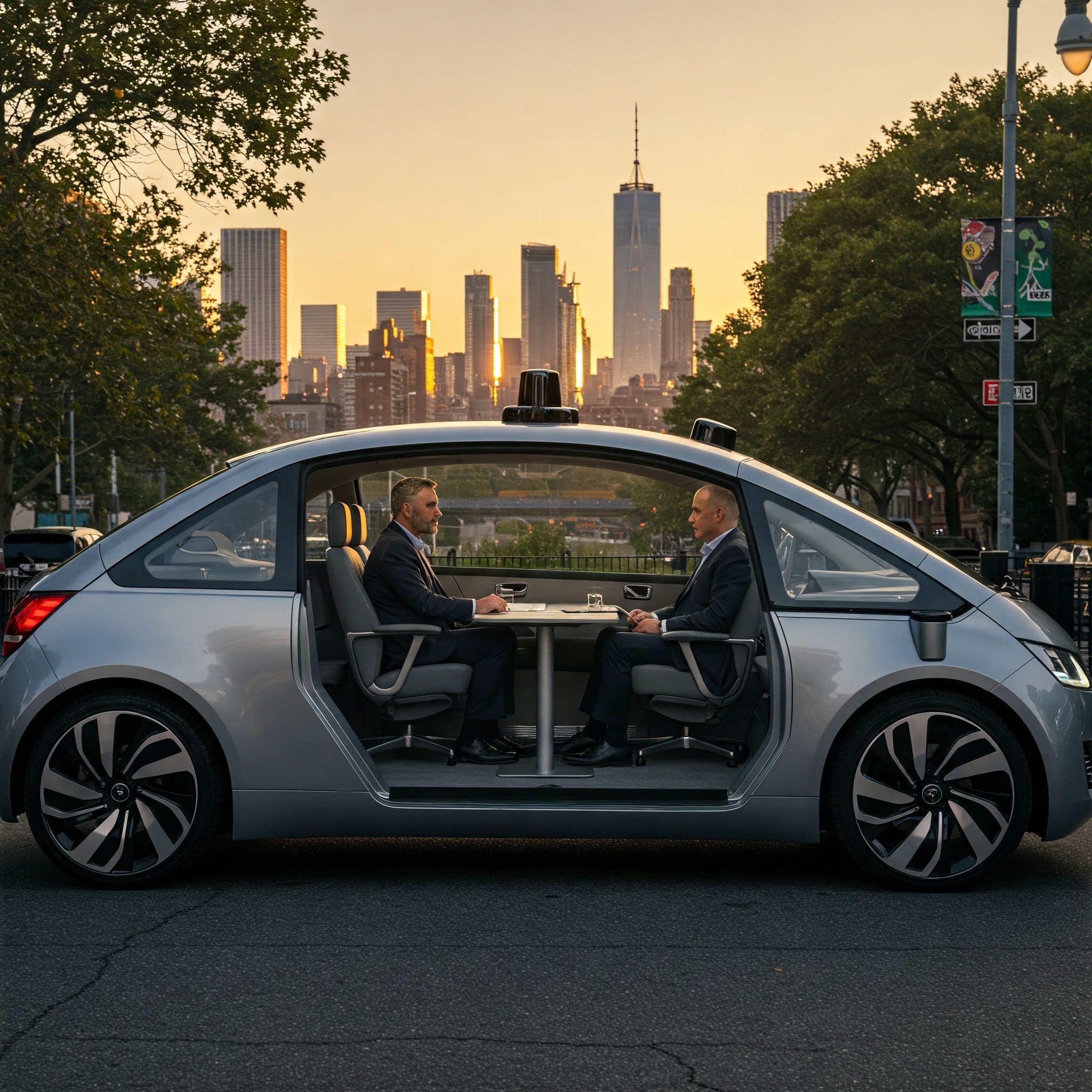}
        \caption{A connected and automated vehicle of SAE level 5}
        \label{fig:gemini1}
    \end{subfigure}%
    \hspace{5mm} 
        \begin{subfigure}[b]{0.48\textwidth}
        \includegraphics[width=\textwidth]{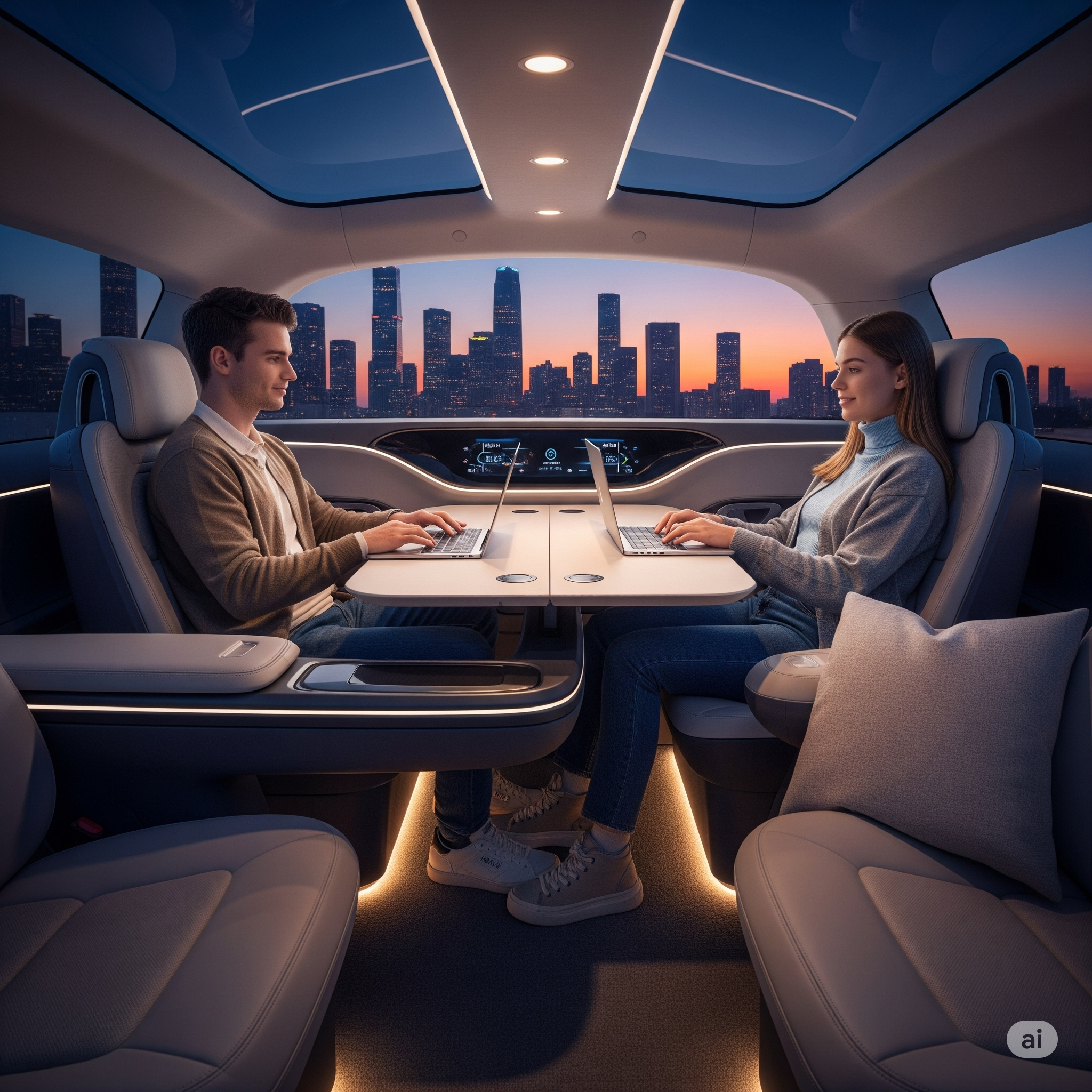}
        \caption{A connected and automated vehicle of SAE level 5}
        \label{fig:gemini2}
    \end{subfigure}%
    \caption{Connected and automated vehicles of SAE level 5 where no human attention is required.}
    \label{geminis}
    \vspace{-15pt}
    \end{figure*}

\section{Existing Technologies in Autonomous Vehicles}
Recently, numerous technologies have made substantial progress, with several autonomous systems already operating on public roads. Most modern vehicles now include advanced driver assistance systems such as adaptive cruise control, lane keeping, and emergency braking, functions that fall under SAE levels 1 and 2; see Fig \ref{fig:SAe LEVELS}. Additionally, certain vehicles exhibit more advanced automation capabilities. Several companies have recently developed systems that allow even self-driving mode (SAE levels 3 and 4). Some of them rely primarily on a camera-only perception system, using deep learning models to interpret the driving environment. A complete self-driving software employs neural networks to detect lanes, vehicles, and obstacles and uses learned planning modules combined with classical control techniques (such as model predictive control (MPC)) to execute maneuvers. In contrast, other companies adopt a sensor-rich approach, integrating LiDAR sensors, radar, and cameras with high-definition maps. This architecture follows a modular design, separating perception, prediction, planning, and control. Such vehicles use MPC for trajectory tracking and depend heavily on rule-based logic and real-time behavioral prediction to navigate safely in urban environments. Although full autonomy (SAE level 5) has not yet been achieved, SAE level 4 systems like robotaxis already operate in selected cities and reflect meaningful progress in deploying autonomous technologies. Existing robo-taxis are classified as SAE level 4 since they operate autonomously within geofenced areas without human input. However, they fall short of level 5 because they cannot function in all driving conditions or locations without constraints.

\vspace{-5pt}
\subsection{Chellenges towards full automation}
There are noticeable attempts to design and develop vehicles that reach higher standards of autonomy. However, the path toward SAE level 5 is far from straightforward. At this level, no human attention or intervention is required at any time, at any location, at any condition, which implies that the vehicle may not even need a steering wheel or driver seat. To get an idea of a vehicle in this category, we refer to Fig. \ref{fig:gemini1} and Fig. \ref{fig:gemini2}, where the passengers are not required to pay attention to traffic conditions. The only necessary interaction is for the users to communicate a destination and preferences such as desired arrival time, music, or cabin temperature. 

Among the most significant barriers to achieving full autonomy is the challenge of safely coordinating CAVs with HDVs on public roads. Unlike CAVs, human drivers can behave in unpredictable ways that are difficult to model and anticipate. This unpredictability limits the ability of even the most advanced decision-making and collision-avoidance algorithms to ensure safety in all situations. Beyond mixed traffic, other critical obstacles include adverse weather conditions, cybersecurity vulnerabilities, and the lack of clear legal and infrastructural support. These factors collectively slow down progress toward full automation. As a result, we are still far from a future where passengers can be told, “You never have to look at the road again.” In fact, some studies suggest that full automation may not be achieved before 2060.

\vspace{-5pt}
\section{Emerging technologies in Mixed Traffic}
Given that full automation is still far from being realized, researchers have focused on developing control algorithms for CAVs operating in mixed-traffic environments encompassing both CAVs and HDVs. Recent works have tackled several challenging problems by modeling and predicting human driving behavior using model or learning-based approaches. These models enable autonomous controllers to adapt in real time based on estimated human intentions.

\begin{figure*}
    \centering
    \begin{subfigure}[b]{0.50\textwidth}
        \includegraphics[width=\textwidth]{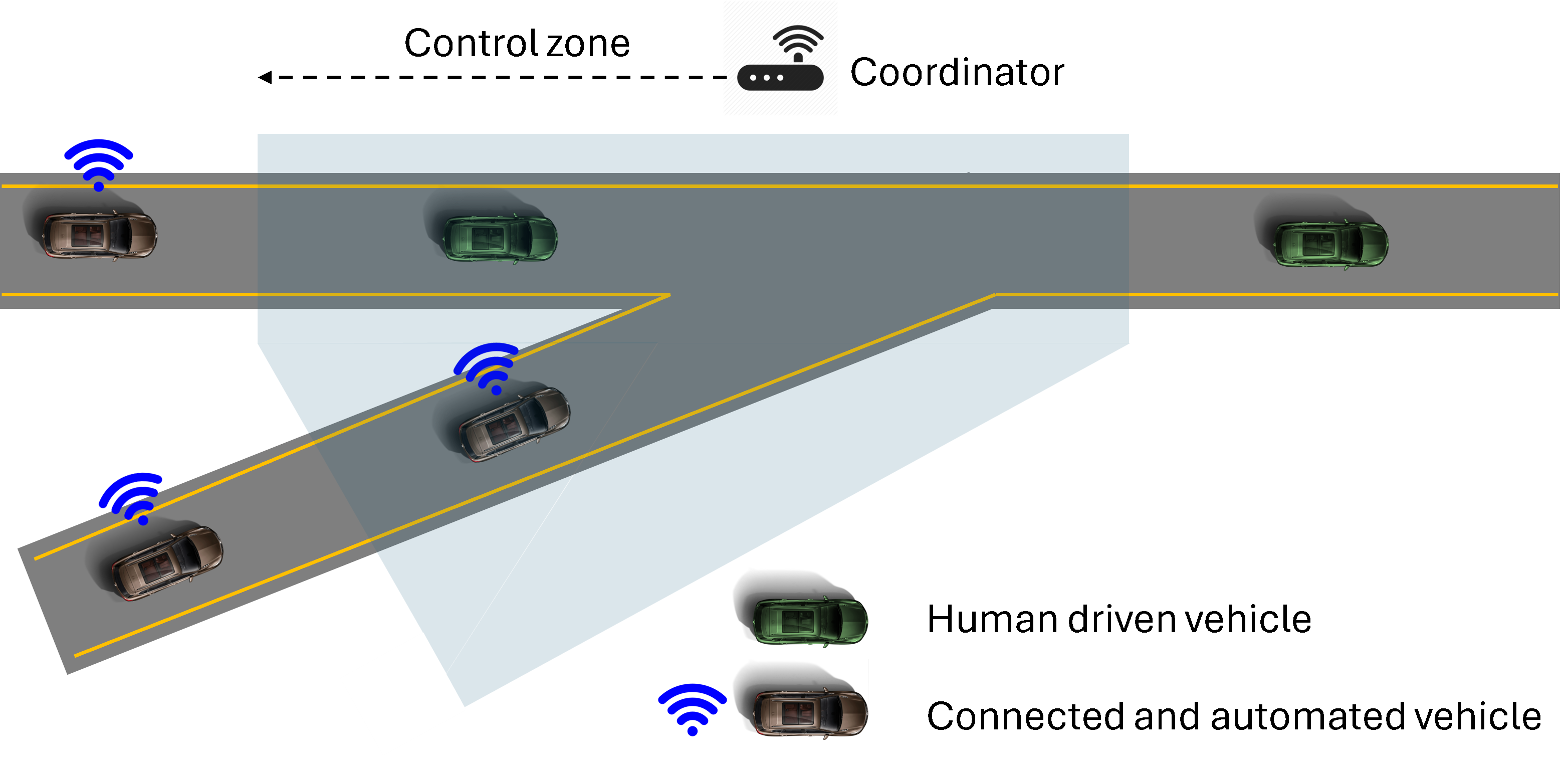}
        \caption{Merging scenario in mixed traffic}
        \label{fig:merging}
    \end{subfigure}%
    \hspace{1mm} 
        \begin{subfigure}[b]{0.48\textwidth}
        \includegraphics[width=\textwidth]{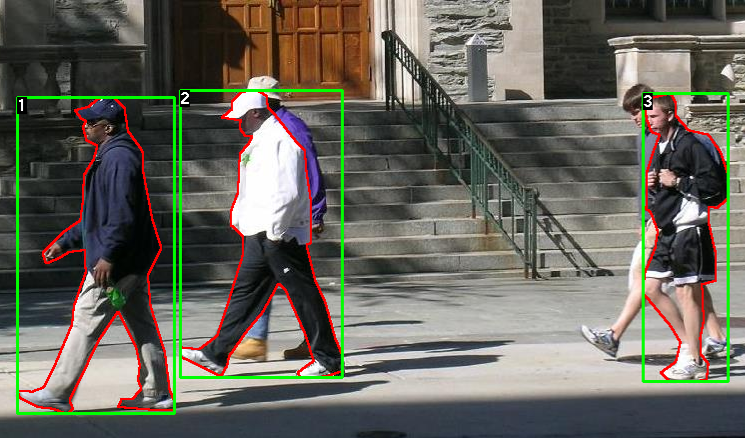}
        \caption{Pedestrian detection}
        \label{fig:pedestrian_detection}
    \end{subfigure}%
    \caption{Scenarios where connected and automated vehicles interfere with humans.}
    \vspace{-12pt}
    \end{figure*}

To illustrate this, we present a high-level idea of a control framework from the literature in a mixed-traffic environment. Consider a merging scenario as depicted in Fig. \ref{fig:merging}. In this example, CAVs and HDVs on two different paths are approaching a merging section where the two paths intersect. The goal here is to efficiently coordinate the trajectories of CAVs while ensuring safety in the presence of HDVs. To achieve this, a coordinator collects data regarding the behavior of HDVs and the planned trajectories of CAVs. The information associated with HDVs can be gathered directly from the CAVs that monitor their surrounding HDVs, or through infrastructure-based sensors such as loop detectors. Interestingly, some studies have also explored the use of drones as mobile data collectors that monitor the transportation network and communicate with CAVs, offering a flexible alternative to static roadside infrastructure.  The coordinator and the CAVs can exchange information within a designated area known as the control zone. Within this zone, CAVs have access to data concerning both other CAVs and nearby HDVs. Using the HDV data, CAVs can predict the future acceleration, speed, and position profiles of the HDVs. Based on these predictions, they can plan energy-efficient and time-optimal trajectories that maintain safety. If HDVs deviate from their predicted trajectory by more than a predefined threshold, the CAVs adapt their trajectories in real time to account for any unexpected human actions. Similar control strategies can be extended to other scenarios, including lane changing, highway driving, and navigating roundabouts.

In parallel, researchers have also studied the interaction between CAVs and pedestrians. Since pedestrian movement can significantly influence traffic flow in urban environments, models are designed to account for their presence and adjust traffic signal timing to improve overall traffic efficiency. Moreover, in cases where pedestrians violate traffic rules and unpredictably enter the roadway, learning-based techniques can again be employed to infer pedestrian intent. For instance, as shown in Fig. \ref{fig:pedestrian_detection}, a CAV performing a turning maneuver detects pedestrians and must simultaneously assess the pedestrian’s intentions and possibly adjust its trajectory accordingly.

Finally, one area that has attracted significant attention over the past decade is the simultaneous control of CAVs and traffic signal phases at signalized intersections. The objective in such problems is to efficiently regulate the traffic light phases based on the traffic flow in each lane, while simultaneously planning the trajectories of the CAVs. The overarching goals are to maximize throughput at the intersection and minimize energy consumption for the CAVs, and indirectly, for the surrounding HDVs as well. This is a particularly challenging problem, as it involves decision variables that are interdependent through coupling constraints. By coupling constraints, we refer to constraints that link the behavior of different subsystems, such as the timing of the traffic signals and the trajectories of the CAVs, in a way that makes them jointly dependent but difficult to optimize independently. Current research addresses this problem through several approaches. One class of methods relies on reinforcement learning, where both the CAVs and the traffic light system operate as agents. Then the goal is to make these agents learn to operate in a specific way that maximizes a predefined reward. Another common approach uses bi-level optimization, in which CAV trajectories are computed in response to traffic signal phases determined by a higher-level optimization problem. A third approach involves joint optimization, where both the actions of the CAVs and the timing of the traffic signals are optimized simultaneously within a single unified problem. It is worth noting that the latter approach can lead to significant computational challenges, particularly as the number of vehicles at the intersection increases.

\vspace{-5pt}
\section{Emerging technologies in full autonomy}
Although full automation (SAE level 5) is still a long way off, numerous studies have demonstrated the potential benefits of a 100\% penetration rate of SAE level 5 CAVs in traffic networks. As a simple example, consider that in a fully automated and connected environment, the need for traditional traffic signals could be eliminated, as CAVs would be capable of coordinating with each other and optimally planning their trajectories with respect to both safety and efficiency. Especially, research has shown that CAVs operating at unsignalized intersections can potentially double the capacity of those intersections compared to traditional setups involving signalized control and HDVs. Communication between CAVs and infrastructure enables optimal control strategies that reduce stop-and-go behavior, thereby decreasing energy consumption and improving overall travel time efficiency. To visualize this, we refer interested readers to the following video, which illustrates the potential of 100\% CAV penetration at an unsignalized intersection: \url{https://youtu.be/Lukwt-L_bhw}.

To fully understand the significance of a fully autonomous transportation network, we can look beyond improvements in safety and efficiency and also consider the potential benefits of accessibility. Aside from enhancing traffic flow and reducing collisions, CAVs can enable more responsive mobility with respect to essential services. For instance, imagine a resident in a congested urban neighborhood who needs to reach the nearest hospital during an emergency. In a CAV-enabled network, this emergency could be automatically communicated to the system, allowing nearby CAVs to adjust their routes in real time. Such coordinated behavior could prioritize the emergency vehicle’s path, minimizing delays and improving access to essential services when time is critical. Nowadays, researchers focus on developing models to improve the fairness and accessibility of transportation networks.

Along these lines, current research has also begun exploring the integration of Large Language Models (LLMs) into CAVs. This integration primarily targets the comfort and personalization aspects of mobility, allowing passengers to communicate naturally with the vehicle. By interpreting spoken or written prompts, LLMs can help the vehicle adapt its driving behavior to passenger preferences. For example, a passenger might enter the vehicle and say, “I'm feeling a bit dizzy today.” An onboard LLM could interpret this as a request for smoother, slower driving and adjust the control parameters accordingly. This type of human-centered interaction has the potential to enhance the travel experience and make CAVs more accessible to a broader population.

Finally, some recent research efforts have proposed lane-free roads in scenarios with a 100\% penetration rate of SAE level 5 CAVs. The idea is that, since the vehicles are fully connected and autonomous, there is no need to adhere to lane boundaries, which can significantly decrease road capacity.
\begin{figure*}
    \centering
    \begin{subfigure}[b]{0.237\textwidth}
        \includegraphics[width=\textwidth]{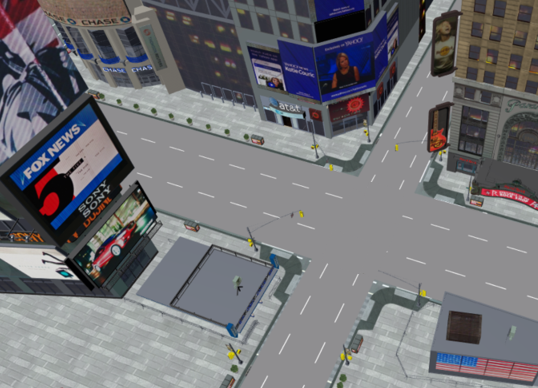}
        \caption{Vissim simulator}
        \label{fig:vissim}
    \end{subfigure}%
    \hspace{1mm} 
        \begin{subfigure}[b]{0.37\textwidth}
        \includegraphics[width=\textwidth]{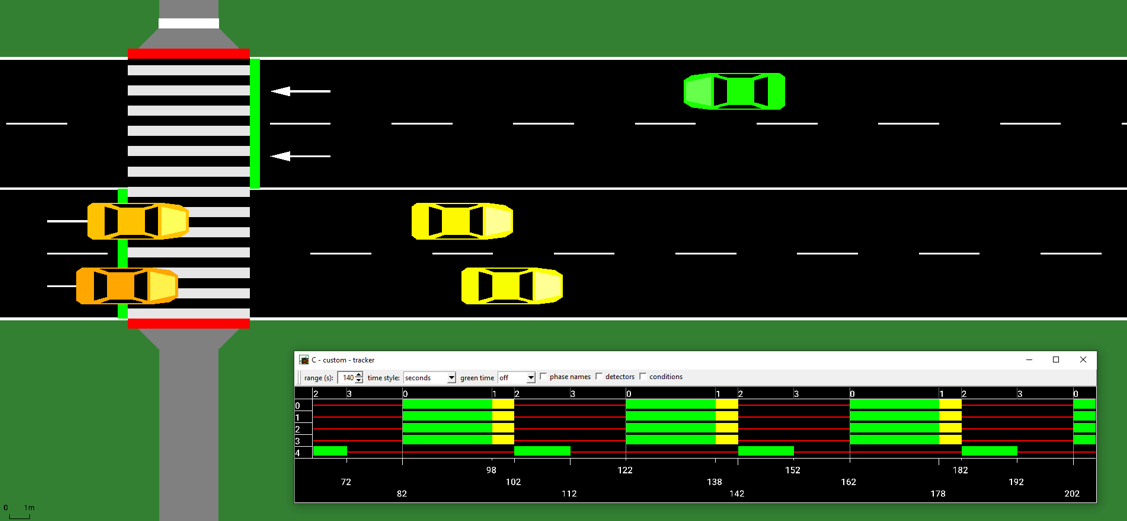}
        \caption{Sumo simulator}
        \label{fig:sumo}
    \end{subfigure}%
        \hspace{1mm} 
        \begin{subfigure}[b]{0.342\textwidth}
        \includegraphics[width=\textwidth]{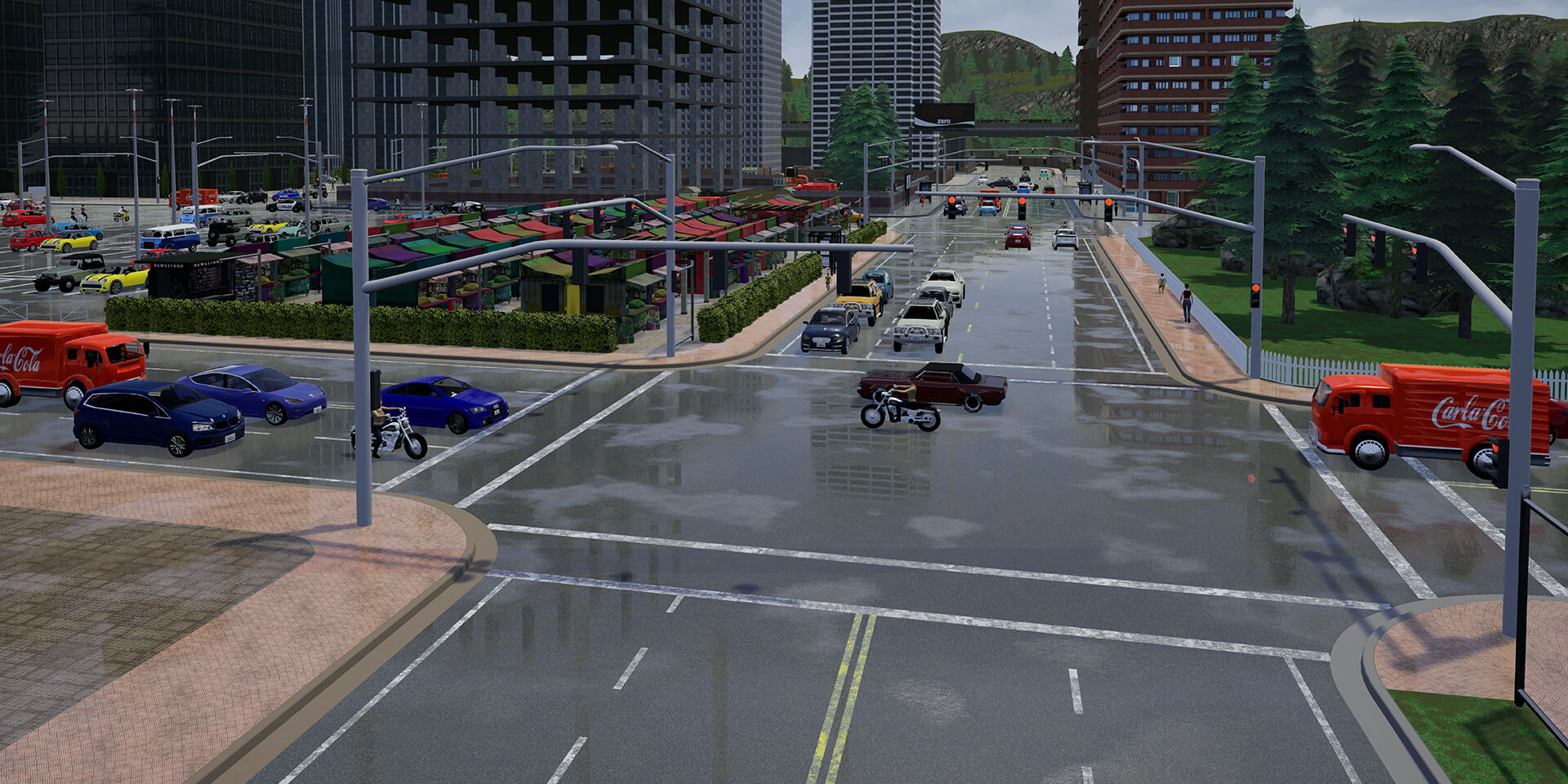}
        \caption{Carla simulator}
        \label{fig:carla}
    \end{subfigure}%
    \caption{Different microsimulation software tools.}
    \vspace{-12pt}
    \end{figure*}

\subsection{Security concerns in full autonomy}
The path toward full autonomy must account for a wide range of safety and security concerns, including those related to information sharing and cybersecurity. A fully connected and autonomous transportation network inherently becomes a potential target for cyberattacks, which could compromise both safety and system reliability. As a result, cybersecurity has become a critical area of research within the CAV ecosystem. Ongoing efforts focus on identifying vulnerabilities, setting safety thresholds, and developing robust protocols to ensure secure communication between vehicles, infrastructure, and users. Addressing these challenges is essential for building public trust and ensuring the safe deployment of autonomous vehicle systems at scale. 

\section{Conducting research on connected and autonomous vehicles}
Several laboratories in industry and academia are actively conducting research on CAVs under all the traffic scenarios already discussed. However, unlike other domains where experiments can be implemented and tested directly within a lab environment, CAV research faces significant challenges when it comes to testing and evaluation under real-world traffic conditions. For that reason, researchers evaluate their algorithms utilizing software or hardware that tries to mimic real-world conditions. Next, we discuss such techniques.

\subsection{Microsimulation software}
Microsimulation software platforms allow researchers to simulate traffic conditions and implement algorithms for CAVs to evaluate their performance on computers. Popular software tools such as SUMO, VISSIM, and CARLA, each offer unique advantages in modeling different aspects of traffic scenarios, from large-scale urban networks to high-fidelity vehicle dynamics; see Figs. \ref{fig:vissim}, \ref{fig:sumo}, \ref{fig:carla}. In recent years, MathWorks has also played a significant role in this domain by offering a suite of tools, including the Autonomous Driving Toolbox and RoadRunner; see Fig. \ref{MathWorks Figures}. These tools enable the implementation and testing of CAV algorithms while integrating seamlessly with other MATLAB tools for LiDAR, camera, radar, and V2X communication, key components of real-world autonomous systems.

\subsection{Scaled cities}
\begin{figure*}
    \centering
    \includegraphics[width=1\linewidth]{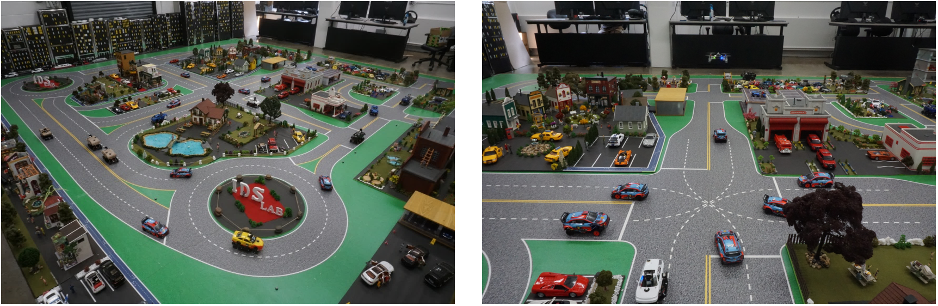}
    \caption{The Information and Decision Science Lab scaled smart city (IDS3C).}
    \label{fig:scaled city}
    \vspace{-14pt}
\end{figure*}

In recent years, several research labs have developed scaled testbeds that replicate real-world traffic conditions in a controlled environment. These setups allow researchers to implement their algorithms on physical hardware and evaluate performance under realistic conditions. Unlike pure simulation, hardware-based experiments introduce practical challenges such as communication delays, localization errors, and tracking limitations, which must be addressed during system design.
One example is the IDS Lab’s scaled city at Cornell University, known as IDS3C; see Fig. \ref{fig:scaled city}. The facility spans 400 square feet and includes 70 robotic cars at a 1:25 scale and 12 quadcopters. IDS3C can emulate complex traffic scenarios, enabling researchers to study the impact of coordination strategies on energy efficiency, throughput, and safety. A VICON motion capture system provides accurate localization, while a central mainframe computer (Intel Xeon w9-3475X) computes the desired trajectories. These are transmitted via UDP/IP to Raspberry Pi units onboard the vehicles. The platform supports experiments in various CAV coordination tasks, including unsignalized intersections, highway platooning, and merging. The quadcopters, known as Crazyflies, are used to investigate aerial-ground interaction, such as last-mile delivery and formation control.
\vspace{-5pt}
\subsection{Virtual Reality Enviroments}
When it comes to research involving mixed traffic environments, the interaction with real human drivers is critically important. Although microsimulation software can include human driver models, researchers still need to validate their algorithms under real driving conditions with actual human participants. Additionally, collecting data from human drivers under different CAV control strategies is essential for evaluating the overall safety of the network. To address this need, virtual reality (VR) setups have been developed that closely mimic real-world conditions. For example, the VR testbed developed in the IDS Lab offers such an alternative by integrating the Meta Quest Pro VR headset with the CARLA traffic simulator. Such an integration enables API-based communication for real-time control, head-tracking response, and interactive scenario customization. Also, this testbed offers a realistic driving experience but also supports the integration of LLMs, enabling researchers to implement and evaluate LLM-based algorithms within the environment.

\subsection{Real-world tests}
Several laboratories have implemented and tested their co-\newpage \noindent ntrol algorithms in real-world settings. Dedicated test tracks have even been constructed for this purpose. One notable example is \textit{MCity}, the University of Michigan’s 32-acre mock city. Opened in 2015, it includes urban and suburban streets, tunnels, overpasses, and simulated pedestrians to help researchers evaluate CAV algorithms under realistic conditions.

More recently, \textit{UC Berkeley} conducted what is believed to be the largest open-road CAV field experiment to date: deploying 100 vehicles equipped with reinforcement-learning–based cruise controllers on a highway. Their goal was to dampen “stop-and-go” traffic waves (also known as phantom jams) and they reported smoother flow, reduced congestion, and lower energy consumption for all drivers.

\begin{figure*}
    \centering
    \begin{subfigure}[b]{0.465\textwidth}
        \includegraphics[width=\textwidth]{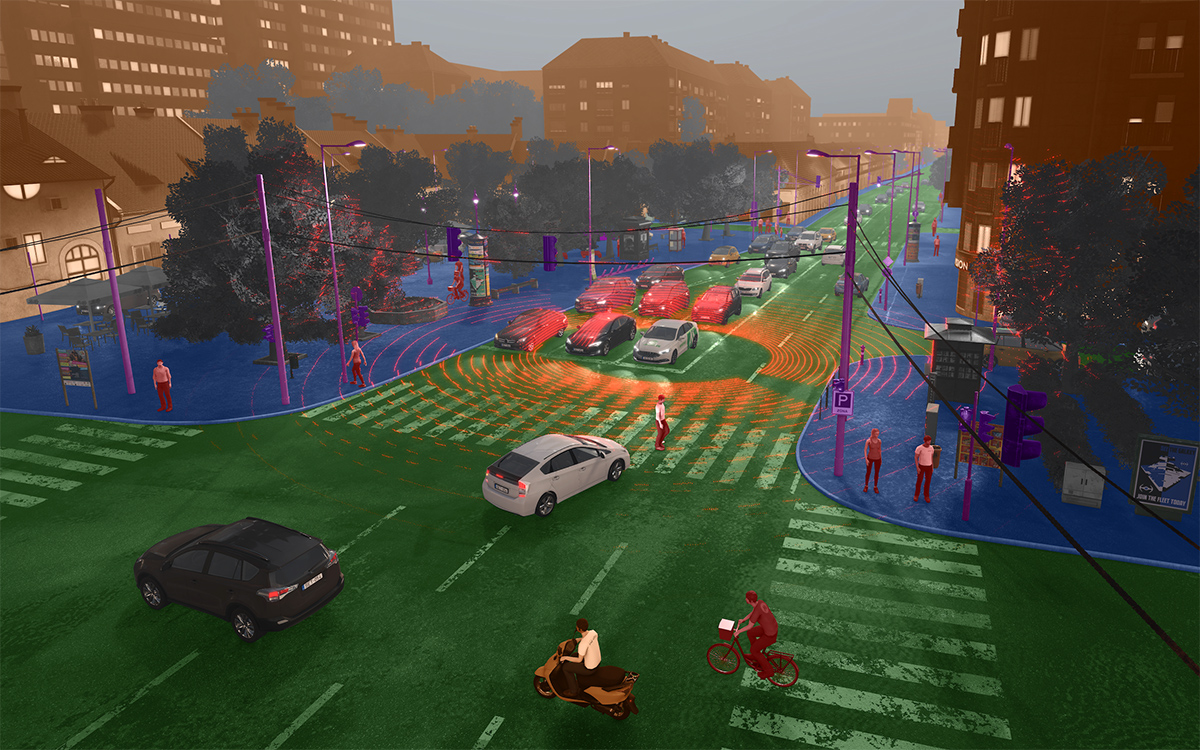}
        \caption{Roadrunner}
        \label{fig:Roadrunner}
    \end{subfigure}%
    \hspace{1mm} 
        \begin{subfigure}[b]{0.517\textwidth}
        \includegraphics[width=\textwidth]{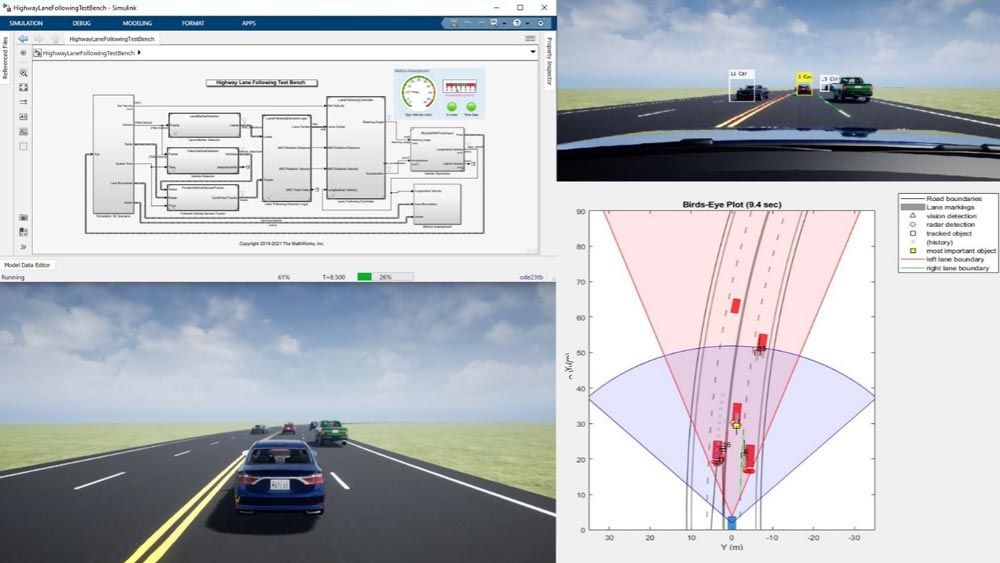}
        \caption{Automated Driving Toolbox with Unreal Engine graphics}
        \label{fig:automated toolbox}
    \end{subfigure}%
    \caption{Simulation tools provided by MathWorks.}
    \label{MathWorks Figures}
    \vspace{-12pt}
    \end{figure*}

\section{Modeling a CAV with PID Speed Control}

In this section, we explore some basic principles of a Proportional-Integral-Derivative (PID) controller and give a tutorial on how we can control the speed of a CAV to a desired speed. PID controller is a widely used feedback control strategy that aims to minimize the error between a desired reference value and the system's current output. The proportional term reacts to the current error, providing a correction that is directly scaled by how far the system is from the target. The integral term accounts for the accumulation of past errors, helping eliminate steady-state offset. The derivative term predicts future error based on the current rate of change, adding damping and improving response stability. Together, these three components enable the controller to react effectively to both immediate deviations and longer-term trends in system behavior.

We consider a CAV that operates in one-dimensional motion (longitudinal direction). The vehicle is modeled as a point mass subject to double integrator dynamics:
\begin{equation}
    \dot{x}(t) = v(t), \quad \dot{v}(t) = u(t)
\end{equation}
where $x(t)$ is the position, $v(t)$ is the velocity, and $u(t)$ is the control input, interpreted as longitudinal acceleration.

The objective is to regulate the vehicle's velocity to a desired reference $v_{\text{ref}}$ using a PID controller. Let the velocity tracking error be defined as:
\newpage
\begin{equation}
    e(t) = v_{\text{ref}} - v(t)
\end{equation}
The control input is computed using the standard PID formulation:
\begin{equation}
    u(t) = K_p e(t) + K_i \int_{0}^{t} e(\tau) d\tau + K_d \frac{de(t)}{dt}
\end{equation}
where $K_p$, $K_i$, and $K_d$ are the proportional, integral, and derivative gains, respectively.

The PID controller aims to minimize the tracking error $e(t)$ by adjusting the acceleration command $u(t)$ such that the vehicle reaches and maintains the desired velocity $v_{\text{ref}}$. This basic control structure provides a useful introduction to CAV behavior under simple closed-loop speed regulation. To experiment with the implementation described above, the reader can refer to the related GitHub repository at the following link: \url{https://github.com/ftzortzo/CAV-Speed-Control-PID}, which provides step-by-step instructions and the necessary MATLAB code.

\section{Conclusion}

CAVs have the potential to fundamentally reshape the future of transportation systems by improving safety, reducing emissions, minimizing delays, and enhancing accessibility. In this article, we provided a high-level overview of the current landscape in CAV technology, including existing deployments, research challenges, and ongoing innovations. We discussed the complexity of mixed traffic environments and the technical hurdles that remain in achieving full autonomy, particularly concerning human unpredictability and cybersecurity. 

To bridge the gap between theoretical development and practical validation, we highlighted various platforms—ranging from microsimulation software to scaled cities and virtual reality environments—that support the evaluation of CAV control strategies under realistic conditions. Lastly, we introduced a simple PID-based controller as an accessible entry point for students interested in modeling CAV behavior.

As the CAV ecosystem continues to evolve, multidisciplinary collaboration between control theory, artificial intelligence, human factors, and policy will be critical. We hope this article serves as an entry point for early-career researchers to engage with this growing and impactful field.

\section*{Acknowledgments}
The authors would like to acknowledge the sources of the figures used in this work. Fig. \ref{fig:SAe LEVELS} was sourced from \url{https://www.sae.org/blog/sae-j3016-update}. Figs. \ref{fig:gemini1} and \ref{fig:gemini2} were generated using Google Gemini. Figs. \ref{fig:Roadrunner} and \ref{fig:automated toolbox} were provided on the MathWorks website \url{https://www.mathworks.com} and Fig. \ref{fig:pedestrian_detection} was obtained from \url{https://www.cis.upenn.edu}.

\begin{figure}
    \centering
    \includegraphics[width=0.994\linewidth]{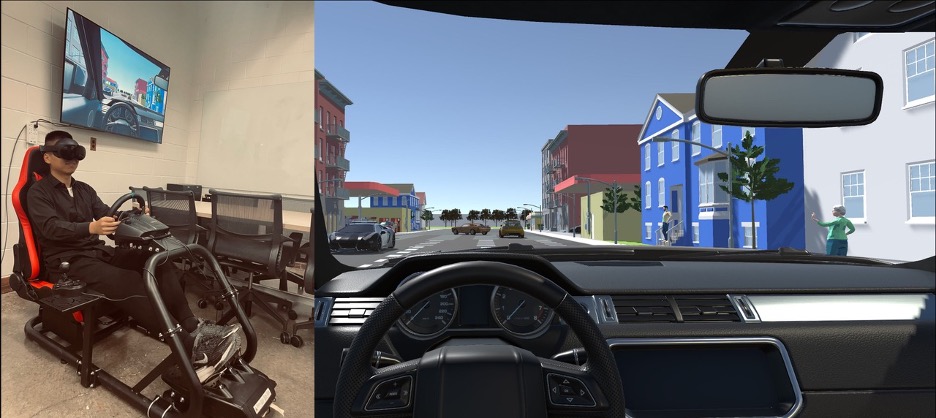}
    \caption{Virtual reality platform.}
    \label{fig:VR}
    \vspace{-5pt}
\end{figure}

\bibliographystyle{IEEEtran}
\nocite{*}
\bibliography{references}

\begin{IEEEbiographynophoto}{Filippos N. Tzortzoglou}
received the Diploma with an integrated M.S. degree in production engineering and management from the Technical University of Crete, Chania, Greece, in 2022. He is currently pursuing a Ph.D. degree with the Civil and Environmental Engineering Department, at Cornell University, Ithaca, NY, USA. Also, in 2024, he joined MathWorks, Natick, MA, USA, as a Research Intern. His research interests lie in the area of automatic control, with applications in transportation and autonomous vehicles.
\end{IEEEbiographynophoto}
\begin{IEEEbiographynophoto}{Andreas A. Malikopoulos}
received a Diploma in mechanical engineering from the National Technical University of Athens (NTUA), Greece, in 2000. He received M.S. and Ph.D. degrees in mechanical engineering at the University of Michigan, Ann Arbor, Michigan, USA, in 2004 and 2008, respectively. He is a professor at the School of Civil and Environmental Engineering at Cornell University and the director of the Information and Decision Science (IDS) Laboratory. Prior to these appointments, he was the Terri Connor Kelly and John Kelly Career Development Professor in the Department of Mechanical Engineering (2017--2023) and the founding Director of the Sociotechnical Systems Center (2019--2023) at the University of Delaware (UD). Before he joined UD, he was the Alvin M. Weinberg Fellow (2010--2017) in the Energy \& Transportation Science Division at Oak Ridge National Laboratory (ORNL), the Deputy Director of the Urban Dynamics Institute (2014--2017) at ORNL, and a Senior Researcher in General Motors Global Research \& Development (2008--2010). His research spans several fields, including analysis, optimization, and control of cyber-physical systems (CPS); decentralized stochastic systems; stochastic scheduling and resource allocation; and learning in complex systems. His research aims to develop theories and data-driven system approaches at the intersection of learning and control for making CPS able to realize their optimal operation while interacting with their environment. He has been an Associate Editor of the IEEE Transactions on Intelligent Vehicles and IEEE Transactions on Intelligent Transportation Systems from 2017 through 2020. He is currently an Associate Editor of Automatica and IEEE Transactions on Automatic Control, and a Senior Editor of IEEE Transactions on Intelligent Transportation Systems. He is a member of SIAM, AAAS, and a Fellow of the ASME.
\end{IEEEbiographynophoto}

\vfill

\end{document}